\documentclass[12pt, letterpaper, notitlepage]{article} 

\usepackage{amsmath}
\usepackage{amssymb}
\usepackage{graphicx}
\usepackage{subcaption}
\usepackage{textcomp}

\usepackage{algorithm}
\usepackage{algorithmic}

\usepackage{fancyhdr}
\begin{document}

\title{Graph Coloring with Quantum Annealing}
\author{Julia Kwok, Kristen Pudenz}
\date{}
\maketitle

\begin{abstract}
We develop a heuristic graph coloring approximation algorithm that uses the D-Wave 2X as an independent set sampler and evaluate its performance against a fully classical implementation. A randomly generated set of small but hard graph instances serves as our test set. Our performance analysis suggests limited quantum advantage in the hybrid quantum-classical algorithm. The quantum edge holds over multiple metrics and suggests that graph problem applications are a good fit for quantum annealers.
\end{abstract}

\section{Introduction}
\label{intro}

Graph coloring is the assignment of labels - colors - to each element of a graph subject to constraints. In its most recognizable form, it is the assignment of a color to each vertex of a graph such that no two adjacent vertices are of the same color, known as vertex coloring. It maps well to a variety of real-world applications of interest in scheduling \cite{Marx2004}, such as job shop scheduling, aircraft flight scheduling, radio bandwith scheduling, and resource allocation, specifically register allocation  \cite{Chaitin1982}. However, it is computationally hard; determining whether a graph is $k$-colorable for any $k\ge3$ is NP-complete while computing the chromatic number (fewest number of colors needed to color the graph) is NP-hard \cite{Garey1974, Garey1979}. For every $k < 3$, a $k$-coloring of a planar graph exists by the four color theorem \cite{appel1989every}, and it is possible to find such a coloring classically in polynomial time. Although NP-hard optimization problems cannot be solved exactly in polynomial time, some can be approximated in polynomial time up to some constant approximation ratio. For this reason,  in practice it is favorable to use heuristic approaches to efficiently find an optimal approximate solution.

We study a heuristic approach to graph coloring based on quantum annealing (QA), which makes use of controllable quantum hardware to solve optimization problems via energy minimization \cite{Farhi2000, Smelyanskiy2012}. The optimization objective is classical and can be expressed as a quadratic unconstrained binary optimization (QUBO) problem (see Equation \ref{eq_QUBO} for details). The process of finding the solution involves a quantum search of the solution space, which is the native function of the QA processor. Quantum annealing has been applied to solve a variety of computational and application based problems such as satisfiability \cite{douglass2015constructing, pudenz2017quantum}, scheduling \cite{Biswas2017, Tran2016, Venturelli2016}, machine learning \cite{khoshaman2018quantum, mott2017solving}, and materials topology \cite{king2018observation}.

Drawing inspiration from previous work both in classical graph coloring approximation algorithms and combinatorial optimization problems on near-term quantum annealing devices, we formulate and benchmark a quantum-assisted graph coloring approximation algorithm. In Section \ref{heuristic-graph-coloring}, we provide an abstract view of the heuristic algorithm and introduce the metrics we use to assess its performance. Subsections \ref{sec_timing} and \ref{problem-set} detail the timing mechanisms used for both versions of the algorithm and the problem set of small but computationally hard graph instances used for benchmarking, respectively. Section \ref{sec_sampling} details the implementation of both the classical and quantum set sampling subroutine. Finally, results from the quantum algorithm run on the test set with the D-Wave 2X are compared with the classical algorithm and interpreted in Sections \ref{results} and \ref{conclusions}. 

\section{Heuristic Graph Coloring}
\label{heuristic-graph-coloring}

\subsection{Algorithm}
Perhaps the most intuitive classical approach to the graph coloring problem is brute force search: enumeration of all possible candidates to a solution until one is found. Although simple and effective, it is highly inefficient. Exact algorithms employing brute-force search to find a $k$-coloring must consider each of the $k^n$ assignments of $k$ colors to $n$ vertices. As the graph size increases, the solution space and consequently the algorithm's complexity scales exponentially. Although heuristic algorithms have been able to achieve $k$-colorings in time and space $O(2.4423^n)$ and $O(2^nn)$ or even $O(1.3289^n)$ and $O(1.7272^n)$ for 3- and 4-colorability, they remain exponentially complex in time and space \cite{Lawler1976, Bjoerklund2009} \cite{Beigel2005, Fomina}. 

The quantum annealing formulation (or broadly the optimization form) of the exact graph coloring problem faces similar practical limitations. Written as a QUBO penalty function, it takes the form:
\begin{equation}
Q_{GC} = \sum_{i=i}^{n} (1 - \sum^{k}_{c=1}b_{i,c})^2 + \sum_{i,j \in E}\sum^{k}_{c=1}b_{i,c}b_{j,c}
\end{equation}
where each variable $b_{i,c}$ represents a graph vertex $i$ with color $c$ \cite{Dahl2013}. The first term penalizes multiple assignments per vertex and the second term penalizes identically labeled adjacent vertices. Here, the programmer is required to determine \textit{a priori} the number of colors $k$ needed for the target graph $G$, which is itself an NP-hard problem. Furthermore, this formulation is not space efficient; it uses $kn$ qubits for an $n$ vertex graph using $k$ colors, which is a resource cost in qubits that scales linearly with both $k$ and $n$. The number of qubits available is severely limited on current generation quantum processors, making space efficiency critical. To avoid these issues, we will not pursue the exact global approach in our work.

Instead, we study approximation algorithms, which are efficient algorithms that produce approximate solutions to NP-hard optimization problems. Greedy approximation algorithms \cite{Black1998}, follow the problem-solving heuristic of selecting a locally optimal choice at each stage. Greedy graph coloring, specifically, scans the set of vertices and chooses to assign a color to locally optimal subset \cite{Mitchem1976}. One version of this heuristic algorithm colors a graph by scanning the entire graph and assigning a color to one subset of vertices at a time. Pseudocode of this approach is detailed in Algorithm \ref{alg:GreedyGraphColoring}.

\begin{algorithm}
\caption{ Greedy Graph Coloring Approximation} \label{alg:GreedyGraphColoring}
\begin{algorithmic}[1]
\REQUIRE graph $G$, sample number $s$
\ENSURE  graph coloring $K$\\
\COMMENT{$I$ is an independent set of graph $G$}
\WHILE{$G \neq \: \emptyset$ }
\STATE {
$P \leftarrow$ choose  \{$I_1, ...I_s$\}\\
$m \leftarrow \underset{I \in P}\max{\lvert I \rvert}$ \\
$M \leftarrow \emptyset$\\
\FORALL{ $I \in P$}
\IF{$\lvert I\rvert =  m$ }
\STATE{
$M \leftarrow \{M,I\}$
}
\ENDIF
\ENDFOR \\
randomly pick $I \in M$ \\
$K \leftarrow \{ K, I \}$\\
$G \leftarrow G - I$\\
}
\ENDWHILE
\RETURN{$K$}
\end{algorithmic}
\end{algorithm}

The algorithm performs iterations of subset identification and removal until the graph is exhausted. It starts by choosing $s$ independent sets of $G$ via some sampling mechanism (ours are detailed in Section \ref{sec_sampling}). These independent sets - sets of vertices such that no two share an edge - are all viable candidate solutions from which the algorithm must choose. In this case, it makes a locally optimal choice based on subset size, selecting one of the largest sampled sets. The aim is to chose a maximum independent set (MIS), an independent set of largest possible size for a graph. The set is then assigned a color, added to the coloring $K$, and removed from the graph. The process is repeated until no nodes remain in the graph.

The choice of independent sets sampled and selected at each iteration may profoundly affect the resulting coloring. In order to reach optimal solutions, additional heuristics can be utilized for specific graph types, such as the inclusion/exclusion of specific vertices in a sampled independent set based on properties such as vertex degree. For the purposes of this paper, we do not implement any of these methods.

\subsection{Success Metrics}				
The three metrics we use to evaluate graph coloring algorithm performance are success probability, wall clock time, and time to solution. In classical computer science, algorithm performance is defined by algorithmic efficiency, relating to the number or amount of computational resources used by the algorithm. Without a physical implementation, an algorithm's efficiency is estimated by theoretical analysis. With one, algorithm efficiency can be empirically measured with a series of standard tests and trials (i.e. benchmarking). Typically, metrics of both time and space efficiency are generated. In algorithms which use quantum as well as classical resources, it is unclear how to analyze the algorithm in terms of space, which is typically defined by RAM and other space usage on classical computing resources. Therefore, we limit our comparison to time efficiency, and measure how long the algorithm takes to run as \textit{wall clock time}.

In practice, additional factors that reflect algorithm utility such as accuracy and/or reliability are equally important; an efficient algorithm is unusable if it is not also reliably accurate. Thus, it is important to consider a performance metric which allows us to infer how successful an algorithm is at a task. Derived from algorithm accuracy on a select set of test instances with known solutions, \textit{success probability} measures how well the graph coloring algorithm is able to find a satisfactory coloring. Oftentimes, functional requirements for accuracy and/or reliability affect the efficiency of an algorithm, which motivates our use of a third measure, \textit{time to solution (TTS)}. TTS merges algorithm (time) efficiency with algorithm accuracy, producing a calculated value that reflects the tradeoff between the two. We select these three metrics as a holistic assessment of the algorithm; it should be successful, time efficient, and provide the best performance tradeoff.

We define success probability as the observed or empirical probability that a single run of the algorithm will result in a $k$-coloring or better of a given graph. For each unique graph instance, we apply the algorithm $r$  times. If $r_s$ runs are able to successfully $k$-color the graph, they are considered successes. The success probability for that instance is 

\begin{equation}
p_{success} = \frac{r_s}{r}
\label{success_probability}
\end{equation}

The wallclock time is the real time elapsed during a single run of the algorithm. For every execution of the graph coloring algorithm on a unique instance, we record the time necessary to reach a solution. Averaging over $r$ runs yields 

\begin{equation}
t_{wallclock} = \frac{\sum^{r}_{i=1}t_i }{r}
\label{wallclock_time}
\end{equation}

Time to solution (TTS) for generalized optimization problems is defined as the time to perform $R$ runs, where $R$ is the number of repetitions needed to find the ground state at least once with probability $p_{target}$

\begin{equation}
R =\left[  \frac{log(1-p_{target})}{log(1-p_{success})} \right]
\end{equation}

and where $p_{success}$ is the instance-dependent success probability defined in Equation \ref{success_probability}  \cite{Ronnow2014}.
In this study, we set $p_{target}=0.99$.

\subsection{Timing}
\label{sec_timing}

We have implemented our time measurements straighforwardly, in order to make the clearest possible comparison between the quantum and classical versions of the graph coloring algorithm. In the classical case, a stopwatch timer is programmed to begin when a new run of an instance is started and end when the instance is colored for that run. The time value is then stored and the timer is reset to begin at the next run. The time across all runs of a particular instance is averaged to produce a single wallclock time per instance.

In the quantum algorithm, only the elapsed real time of the classical outer loop is measured via stopwatch timer. The quantum portion of the algorithm is timed and returned separately by the quantum processor. This time includes further classical preparation of the QUBO for the quantum processor by its host server, programming time to set up the optimization on the quantum hardware, and $s$ iterations of a sampling procedure that includes the quantum anneal itself and a small amount of quantum hardware overhead. We add the total aggregate sampling time to the classical stopwatch time to arrive at the total quantum wallclock time. This is a more complete picture than using only the anneal time, common in many quantum annealing problem studies.
 
We discount the network lag to access the quantum system, the processor's small initial setup protocol, and the embedding time. The network lag could be avoided by co-locating the classical processing with the quantum system, and the non problem specific initial setup could be done in parallel with the classical portion of the hybrid algorithm in a more integrated setting.

\begin{figure}
\centering
\includegraphics[width=2.67in]{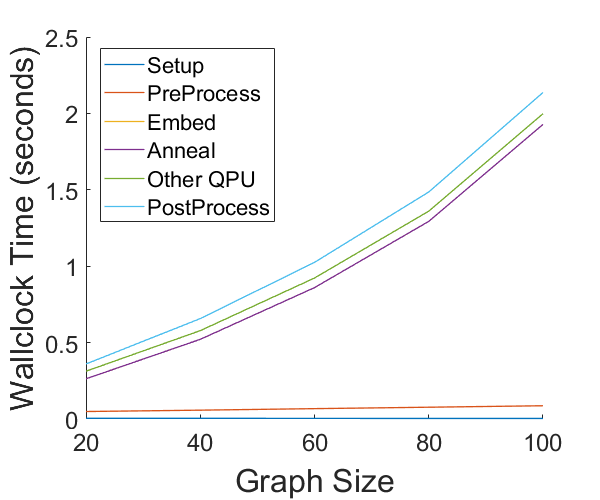}
\includegraphics[width=2.67in]{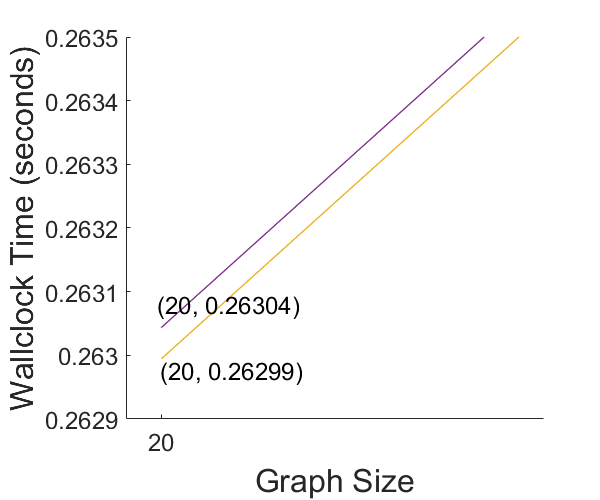}
\caption{Stacked Timing Breakdown: real elapsed time vs graph size. Shown is the averaged time across all $k_{induced}=3$ instances and runs with sample size $s=1$, split into categories (left). The difference in wallclock time between a point on a line and a corresponding point on the line underneath (starting at $t = 0$) is the average time recorded for that time category. On the right is a enlarged and labeled region of the left figure around $n=20$, to highlight the difference in $Embed$ and $Anneal$ times.}
\label{Stacked_Timing}
\end{figure}

Embedding, the problem of matching the connectivity of the graph to be colored to the hardware coupling graph, is a hard problem and a significant contributor to runtime (see Figure \ref{Stacked_Timing}). We leave it out of the time curves in this work because we study problems that are at the limit of what is realizable with the size of quantum processor available to us. A larger processor (or one with higher connectivity) would make the embedding problem easier. Since embedding is not the focus of our work, we leave these problems aside.

\subsection{Problem Set}
\label{problem-set}

We test our algorithm on a set of randomly generated Erdos-Renyi graphs based on previous work on the generation of hard planning problems at the solvable/unsolvable phase transition of NP-complete problems. Rather than modeling graph problems from real-world applications, many of which are too large to be programmed on current quantum processors, we chose to generate small but hard instances. For graph coloring problems, a sharp phase transition in the threshold in the $k$-colorability of graphs has been established for all $k \geq 3$ in terms of the parameter $c = m/n = p \times n$, where $m$ is the number of edges of the graph, $n$ is the number of vertices, and $p$ is the probability with which an edge is formed between any pair of vertices \cite{Achlioptas1999}. The $k$-colorability threshold scales as $c=k\log k$ in the leading term, but the exact location of this threshold remains an open question \cite{CojaOghlan2013}. Experimentally, for 3-color graphs of size $n=10, 12, 14, 16,$ and $18$, the phase transition was observed at approximately $c=4.5$ \cite{Rieffel2014}, so we chose the same value to generate our test set although it includes larger graphs.

To create a robust test set, we targeted problem instances of varying size and chromatic number. Our test set includes instances of graph size $n=20, 40,$ and $60$. We chose a range of graph sizes such that algorithm performance can be readily observed and analyzed, specifically avoiding smaller, previously studied problems as well as problems too large to be programmed onto current quantum annealers. Each $n$ has a corresponding graph connectivity $p$, calculated by setting the parameter $c=4.5$. We induced hidden $3$-colorings, generating twenty unique instances for each graph size $n$, totalling 60 graph instances.

\section{Classical and Quantum Sampling Subroutine}
\label{sec_sampling}
\subsection{Classical}
\label{sec_classical}

To sample adequately large independent sets for the greedy graph coloring algorithm classically, we implemented a subroutine of the best known graph coloring approximation algorithm. This algorithm extends earlier work on approximating maximum independent sets by excluding subgraphs to graph coloring, and returns a coloring of size at most $O(n(\log \log n)^2(\log n)^{-3})$ times the chromatic number $k$ \cite{Bopanna1992, Halldorsson1993}. We use its independent set sampling routine because it is the closest classical equivalent to the quantum sampler, and has one of the best performance guarantees for an approximate graph coloring algorithm.

The base algorithm for approximating independent sets, $Ramsey$, creates an independent set by recursively searching through the neighborhoods $N(v)$ and non-neighborhoods $\overline{N}(v)$ of a chosen vertex $v$, called the pivot node, and returns the largest accumulated result. It is detailed in Algorithm \ref{Ramsey}.

\begin{algorithm}
\caption{Ramsey} \label{Ramsey}
\begin{algorithmic}[1]
\REQUIRE graph $G$
\ENSURE   clique $C$, independent set $I$\\
\IF{$G = \emptyset$}
\RETURN ($\emptyset, \emptyset$)
\ENDIF
\\ \textbf{choose some} $v \in G$
\\ ($C_1, I_1) \leftarrow$ Ramsey($N(v)$)
\\ ($C_2, I_2) \leftarrow$ Ramsey($\overline{N}(v)$)
\RETURN (\textbf{larger of} ($C_1 \cup \{v\}, C_2$), \textbf{larger of} ($I_1, I_2 \cup \{v\}$))
\end{algorithmic}
\end{algorithm}

If the graph does not contain any large cliques (subsets of vertices such that every two are adjacent), $Ramsey$ performs well as a standalone independent set sampling algorithm. However, if the graph does contain large cliques, a performance guarantee on the algorithm cannot be made. The method employed in the $CliqueRemoval$ algorithm provides a solution to this by exhaustively removing the cliques found through $Ramsey$. It then returns the largest of the independent sets found, along with the sequence of cliques found (and subsequently removed), described in Algorithm \ref{CliqueRemoval}.

\begin{algorithm}
\caption{CliqueRemoval} \label{CliqueRemoval}
\begin{algorithmic}[1]
\REQUIRE graph $G$
\ENSURE   independent set $I$, series of cliques $C_1 ... C_n$\\
$i \leftarrow 1$
\\ $(C_i, I_i) \leftarrow$ Ramsey($G$)
\WHILE{$G \neq \emptyset$}
\STATE{
$G \leftarrow G - C_i$
\\ $ i \leftarrow i + 1$
\\ ($C_i, I_i) \leftarrow$ Ramsey($G$)
}
\ENDWHILE
\RETURN (($\max^i_{j=1}I_j$), $\{C_1, C_2, ..., C_i\}$)
\end{algorithmic}
\end{algorithm}

Haldorsson (1993) \cite{Halldorsson1993} further improves upon the accuracy of this earlier algorithm and obtains the aforementioned performance guarantee by incorporating $CliqueRemoval$ into $SampleIS$, shown below in Algorithm \ref{SampleIS}. When $CliqueRemoval$ fails to find a large independent set approximation, the size of the non-neighborhood of the independent set must be large, specifying the need for deeper recursion. $SampleIS$ addresses this observation for $k$-colorable graphs. It starts by randomly choosing a set of nodes, and recursively searches and accumulates nodes in the non-neighborhood of this set, taking advantage of $CliqueRemoval$ to return the largest independent set and removing cliques found in the non-neighborhood.

\begin{algorithm}
\caption{SampleIS} \label{SampleIS}
\begin{algorithmic}[1]
\REQUIRE graph $G$, color $k$
\ENSURE  independent set $I$ of graph $G$ \\
\COMMENT{$G$ is $k$-colorable, $\lvert G \rvert = n$}
\IF{$\lvert G \rvert \leq 1$}
\RETURN $G$
\ENDIF
\WHILE{\TRUE}
\STATE{
randomly pick a set $I$ of $\log_kn$ nodes
\IF{$I \: is \: independent$}
\IF{$\lvert \overline{N}(I) \geq n/k \cdot \log n/2 \log \log n$}
\RETURN ($I \cup$ SampleIS($ \overline{N}(I), k$ )
\ELSE
\STATE{
$I_2 =$ CliqueRemoval($ \overline{N}(I)$) $\cup I$
\IF{$\lvert I_2 \rvert \geq \log^3n/6 \log \log n$}
\RETURN ($I \cup I_2$)
\ENDIF
}
\ENDIF
\ENDIF
}
\ENDWHILE
\end{algorithmic}
\end{algorithm}

 The algorithms were implemented as described in Matlab; no changes were made.

$SampleIS$ returns a single independent set. In order to return the $s$ independent sets of $G$ specified in Algorithm \ref{alg:GreedyGraphColoring},  $SampleIS$ must be called (on graph $G$ with $k$-coloring) $s$ times. 

\begin{figure}
\centering
\includegraphics[width=2.67in]{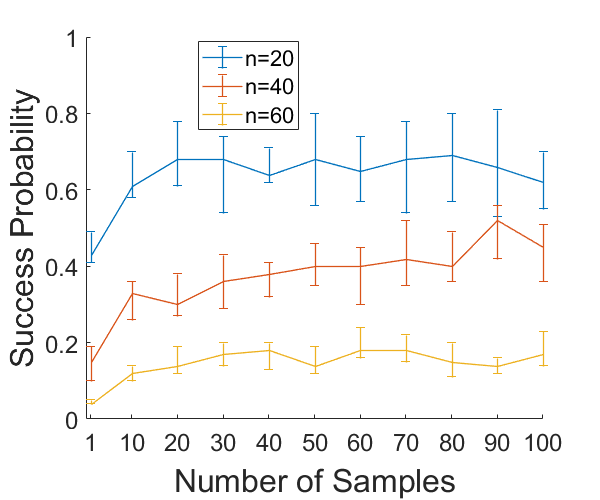}
\includegraphics[width=2.67in]{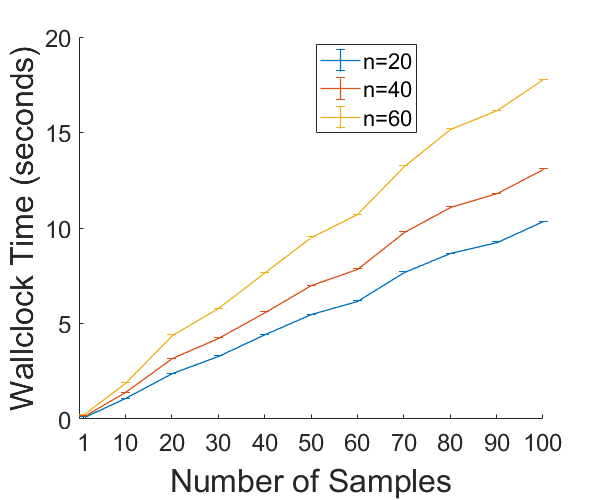}
\includegraphics[width=2.67in]{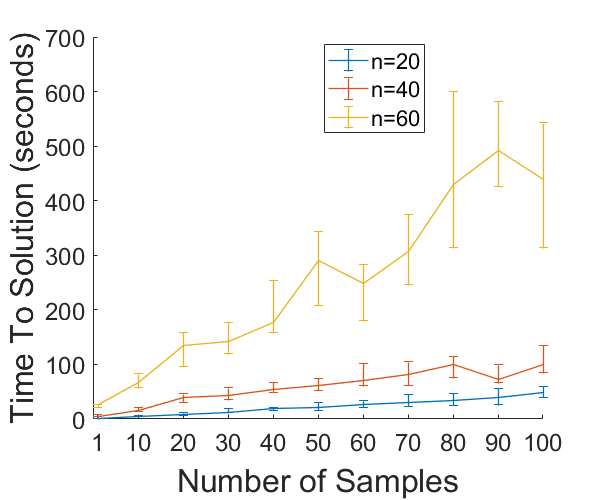}
\caption{Classical Algorithm Results: success probability, wallclock time, and time to solution vs. sample size. Each of the $n_i$ test instances of a specific graph size was run $r$ times, which produced  $n_ir$ data points for success probability and wallclock time. The success probability line plot graphs the median success probability of the data, while the error bar limits represent the 40th and 60th percentile of the data ($\pm 10$ percentile). The wallclock time line plot graphs the median wallclock time of the data, while the error bars represent the standard deviation of the data ($\pm \frac{\sigma}{2}$). The time-to-solution (TTS) line plot graphs the calculated TTS values from the corresponding success probability and wallclock time. Error bars on this plot are calculated from the success probability and wallclock time error bars.}
\label{Classical_Results}
\end{figure}

The classical sampling subroutine has one degree of freedom: the sample number $s$, which specifies the number of independent sets to return from the subroutine. The value of $k$ is fixed for our study because we generate a test set with guaranteed $k$-colorings. To observe the effects of varying the sample number on the performance of the algorithm, we ran the classical graph coloring algorithm with $s = 1$ and then $s = 10, 20, 30 ... 100$ on the test set. This range was chosen conservatively; $s$ was kept within two orders of magnitude of the minimum value in order to observe any trends while minimizing the time cost. The results are shown in Figure \ref{Classical_Results}.

The three graphs in Figure \ref{Classical_Results} collectively show the performance of the classical algorithm relative to sample number for the test set. Although the sucess probability decreases with larger graph sizes, there does not appear to be any obvious trend in the success probability when varying the sample number; it increases from $s= 1$ to $s=10$ and generally plateaus. Predictably, the wallclock time increases linearly with increasing sample number, but at increasing rates for increasing graph sizes. This is reflected in the TTS plot, which also shows an upward trend with an increasing number of samples for all graph sizes. When optimizing the implementation of either algorithm, we seek to to maximize the success probability while minimizing wallclock time, thus minimizing the TTS. In the classical case, the time/success tradeoff is largely influenced by wallclock time. Therefore, the smallest possible sample number $s = 1$ - which yields the shortest wallclock time and TTS - is the optimal choice for the greedy algorithm.

\subsection{Quantum}
The quantum assisted independent set sampling subroutine, similar to its classical counterpart, processes an input graph $G$ and returns a large independent set. To achieve this, the quantum annealer uses a low energy controllable quantum system to minimize a discrete energy function. We formulate the independent set search as a QUBO penalty function suitable for implementation on the DW2 quantum annealing processor. A QUBO is defined by an $N \times N$ upper-triangular matrix $Q$ of real weights as minimizing the function

\begin{equation}
f(x) = \min_{x \in \{ 0, 1 \}^n} x^TQx = \sum_i{Q_{i,i}x_i} + \sum_{i<j}{Q_{i,j}x_ix_j}
\label{eq_QUBO}
\end{equation}

over the binary vector $x$, where the diagonal terms $Q_{i,i}$ are the linear coefficients and the nonzero off-diagonal terms are the quadratic coefficients $Q_{i,j}$. To find suitably large independent sets, we use the optimization form of the maximum independent set problem, expressed in QUBO form as \cite{Calude, Caludea}

\begin{equation}
f_{MIS}(x) = -\sum_{i=1}^n{x_i} + \alpha \sum_{i, j \in E}{x_ix_j} \;\;\;\;\;  \alpha \geq 2,  x\in \{ 0, 1\}
\label{MIS_QUBO}
\end{equation}

The objective of an optimization solver under these constraints is to return the largest subset of nodes in a graph such that no two nodes are adjacent. The first term in the QUBO penalty function maximizes the size of the independent set by rewarding the inclusion of nodes, while the second sets an adjustable penalty $\alpha$ for inclusion of any two nodes connected by an edge. The problem (input) graph, in this case, is the same as the graph programmed into the quantum annealer.

The quantum method of obtaining of obtaining a large independent set approximation differs from the classical method in that there is no scanning of the graph to accumulate vertices. Quantum annealing allows for traversal of the solution space of independent sets to find locally if not globally optimimum solutions. Classically, exploration of the solution space is extremely limited and dependent on how well the initial set of vertices is selected. As the solution space grows, this inherent locality becomes undesirable. 

One anneal (run) of the problem on the quantum annealing processor results in one read - one solution. In order to return the $s$ independent sets of $G$ as requested in Algorithm \ref{alg:GreedyGraphColoring}, the processor must peform $s$ anneals and reads.  In this study we chose to use the standard $20 \mu s$ anneal schedule, so the average anneal time of the MIS problem is on the order of microseconds ($\mu s$). Compared to the deciseconds necessary to obtain a classical sample, the relative time cost of obtaining additional independent set approximations through the quantum processor is much smaller than that of the classical algorithm.

The degrees of freedom in the quantum optimization approach are $\alpha$, introduced in Equation \ref{MIS_QUBO}, and the number of samples $s$ from Algorithm \ref{alg:GreedyGraphColoring}. In implementation, there are additional quantum annealing controls that serve as free variables (such as the anneal schedule), but these are kept at default to avoid introducing more degrees of freedom. To address $\alpha$, we ran the MIS problem on a randomly generated test set with varying values of $\alpha$ to determine what, if any, was the optimal value which produced the largest viable solutions. We concluded that there is no added advantage to increasing the penalty past the minimum value of 2.

\begin{figure}
\centering
\includegraphics[width=2.67in]{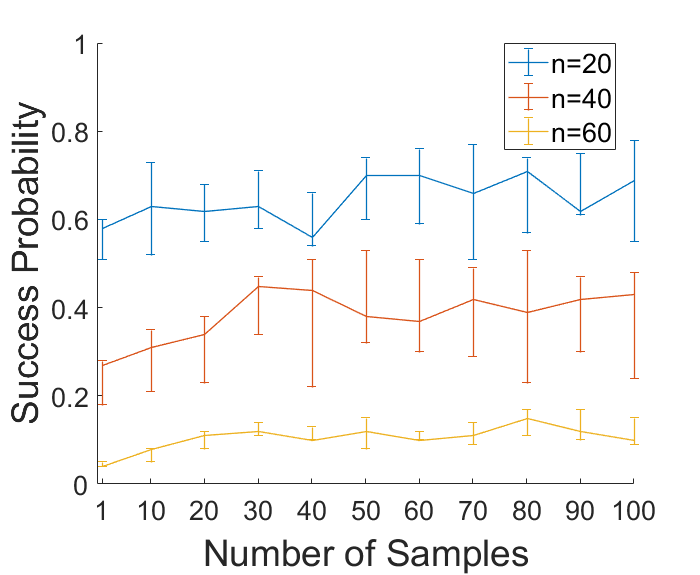}
\includegraphics[width=2.67in]{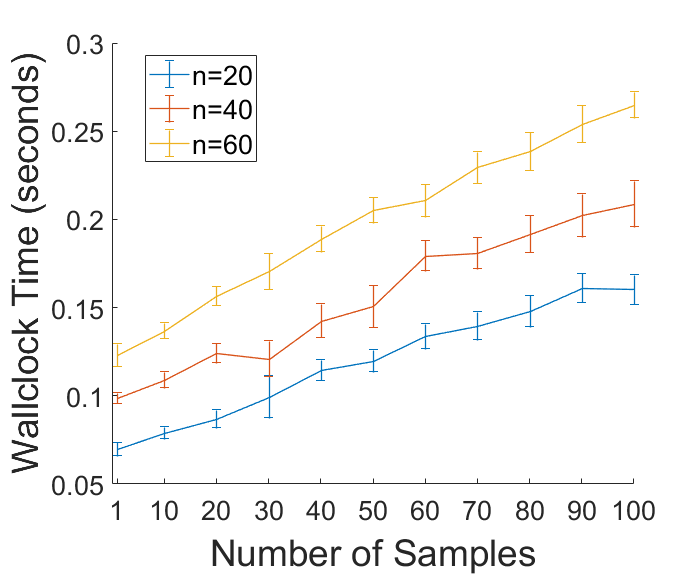}
\includegraphics[width=2.67in]{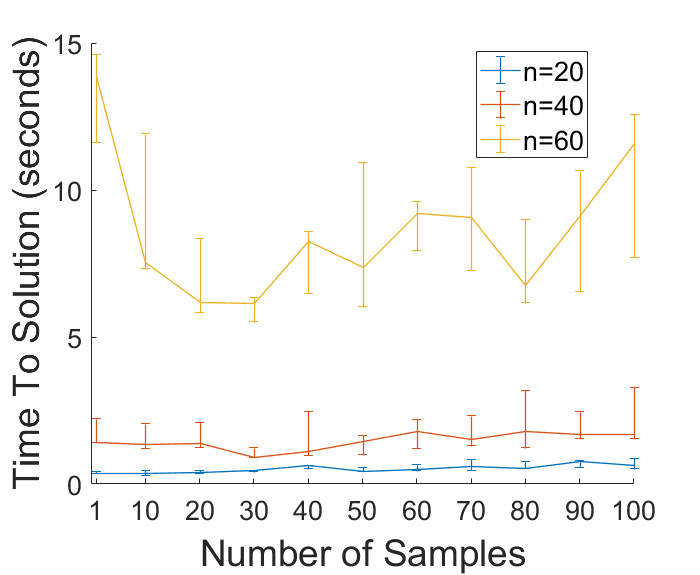}
\caption{Hybrid quantum-classical results: success probability, wallclock time, and time to solution vs. sample size. Each of the $n_i$ test instances of a specific graph size was run $r$ times, which produced  $n_ir$ data points for success probability and wallclock time. The success probability line plot graphs the median success probability of the data, while the error bar limits represent the 40th and 60th percentile of the data ($\pm 10$ percentile). The wallclock time line plot graphs the median wallclock time of the data, while the error bars represent the standard deviation of the data ($\pm \frac{\sigma}{2}$). Note that the wallclock time referred to here includes the classical outer loop of the algorithm as well as the relevant quantum processing time. The time to solution (TTS) line plot graphs the calculated values for TTS and error at each sample number using the corresponding plot points from the success probability and the wallclock time.}
\label{Hybrid_Results}
\end{figure}

In order to observe the effects of varying the sample size on performance, we ran the quantum graph coloring algorithm with $s = 1$ and then $s = 10, 20, 30 ... 100$ on the test set. Although we ran trials with values of $s > 100$, there was no observable improvement in performance for test instances so these results are omitted. The results are shown in Figure \ref{Hybrid_Results}.

The three graphs in Figure \ref{Hybrid_Results} collectively show the performance of the quantum algorithm relative to sample number for the test set. As with the classical implementation, success probability decreases while wallclock time increases with increasing graph size. Similarly, there is also no noticable trend in the success probability as the number of samples increases. The wallclock time plot, as with the classical result, increases as the number of samples increases. However, the rate of increase and order of magnitude of the wallclock time are much smaller in this case. Although it is clear that smaller sample numbers are more time efficient, given the ambiguity of the success plot, we look to the TTS plot to draw further insights. The TTS graph is relatively stagnant for graph sizes $n =20$ and $n = 40$, but shows discernable behaviour at $n =60$. We reason that for smaller sample sizes the algorithm has a smaller success probability but takes less time, and any gains from faster wallclock time are outweighed by the low success probability. For larger sample sizes, the algorithm has a higher success probability but takes more time; the impact of a longer run time exceeds the accuracy advantage. At $s=30$, the curve has reached a sort of minimum at which the increased accuracy of additional samples is traded off equally with an increase in runtime. After this point, the plot does change but does not following an apparent trend, and dwells within this regime until it starts increasing at $s=80$. Therefore, we designate $s=30$ as the optimal number of samples for the quantum graph coloring algorithm.

\section{Results}
\label{results}

\begin{figure}
\centering
\includegraphics[width=2.67in]{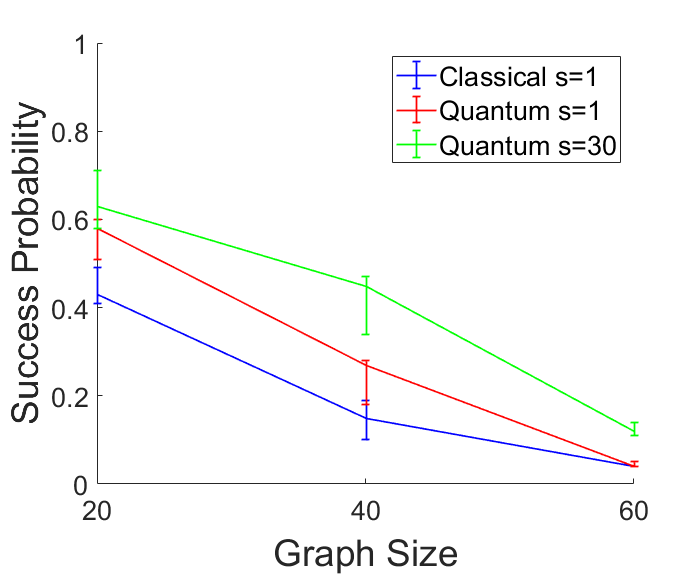}
\includegraphics[width=2.67in]{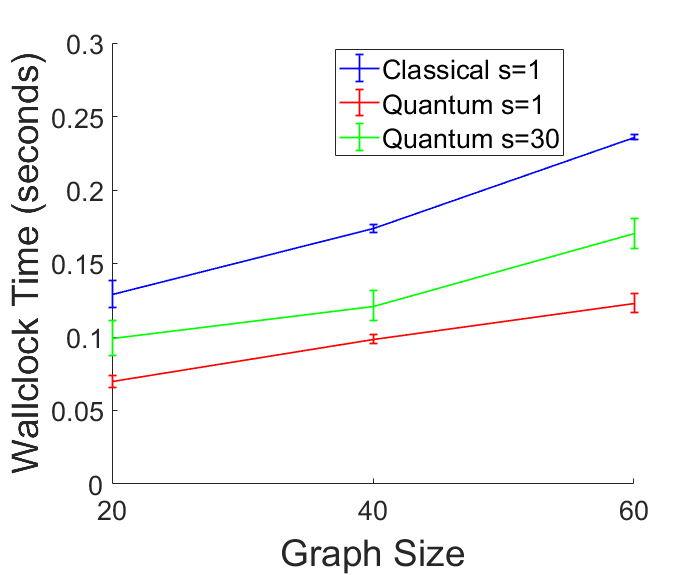}
\includegraphics[width=2.67in]{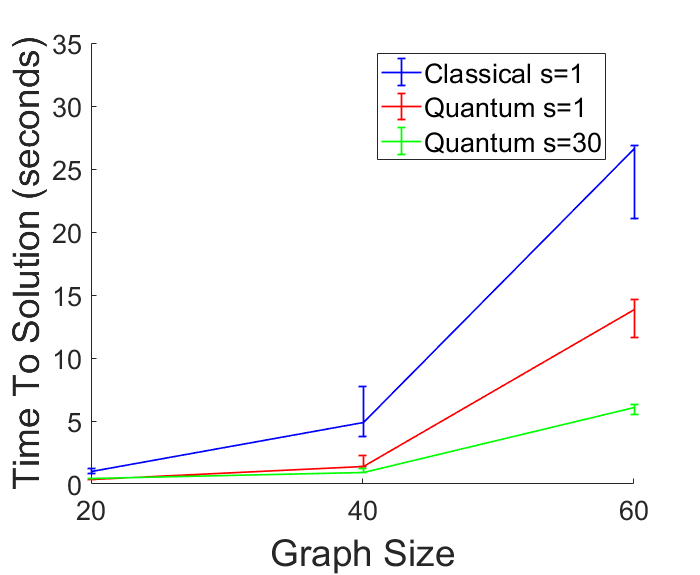}
\caption{Hybrid quantum-classical comparison results: success probability, wallclock time, and time to solution vs graph size. Performance of three optimized graph coloring algorithms as a function of graph size. The median value of the instance success probabilities and wallclock time are plotted on the top left and top right, respectively. Error bar limits represent the 40th and 60th percentile of the data ($\pm 10$ percentile). Time-to-solution (TTS) is plotted at bottom center, using the points from the success probability and the wallclock time at the specified graph sizes. The error bars on this plot are calculated from the corresponding error values for success probability and wallclock time.}
\label{Comparison_Results}
\end{figure}

Performance of the classical and quantum algorithms on our test set is shown in Figure \ref{Comparison_Results}.
For all three algorithms shown, as graph size increases, success probability decreases while wallclock time increases. As a result, the TTS increases follow a nearly exponential curve with respect to graph size. This is not unexpected for an NP-hard problem; the question is whether we can differentiate the quantum performance from the classical.

The classical approximation algorithm coloring guarantee scales with a factor of $n(\log \log n)^2(\log n)^{-3}$, as discussed in Section \ref{sec_classical}. That is, there is less of a guarantee of successful (close to chromatic number) coloring as the graph size increases. The time complexity of the classical algorithm also scales with increasing problem size. As the number of vertices increases, the number of operations needed to scan the graph recursively for large independent sets increases. Furthermore, time devoted to running the classical control loop outside the MIS subroutine also increases for larger problem sizes, and affects both algorithms. It follows that the TTS will also increase.
 
In the quantum case, we see the wallclock time increase at a slightly slower rate than the classical algorithm. Recall that we include the classical setup, preprocessing, and postprocessing times, which scale up with increasing graph size. However, the behavior of the quantum portion of the wallclock time is not as clear. We can describe the quantum time as $s \times (k + r_a ) \times t_{sample}$, where $s$ is the number of samples, $k$ is the number of MIS iterations (graph colors), $r_a$ is the number of sampling attempts (iterations that returned no valid independent sets), and $t_{sample} = t_{anneal} + t_{readout} + t_{delay}$  is the sampling time. For our particular hardware, $t_{anneal} = 20 \mu s$, $t_{readout} = 41 \mu s$, and $t_{delay} = 309 \mu s$, meaning $t_{sample} = 370 \mu s$. Quantum time does not directly depend on graph size, but does depend on $k$ and $r_a$, which are correlated with increased problem hardness and decreased success probability. The effects of embedding large graphs and challenge of an increasing number of local solution minima lead to decreasing success as graph size increases. Larger graphs have higher embedding overhead (more physical qubits per logical qubit) and are more likely to yield physical solutions with "broken" logical qubits that do not decode to an independent set. Larger solution spaces result in increasing chances of selecting local minima that yield smaller independent sets, causing the heuristic to return more colorings with $k>3$. These factors depress success probability for larger graph sizes. 

We see better performance using the quantum annealer with respect to all three metrics. We compare the performance of the classical algorithm with that of the quantum algorithm using $s=1$ and $s=30$, which we refer to as the sub-optimum and optimum quantum algorithms, respectively. We include the $s=1$ quantum algorithm as a direct comparison to the $s=1$ classical algorithm. The success probability of the classical algorithm starts lower than that of the two quantum algorithms for $n=20$ and $40$, only equaling the performance of the sub-optimum quantum algorithm at $n=60$. With respect to wallclock time, the classical algorithm takes significantly longer to return a graph coloring for all graph sizes compared to the quantum algorithms. The same is true of the classical algorithm's time-to-solution performance. The rate at which the TTS of the classical algorithm increases as the graph size increases is faster than that of both quantum algorithms. Both the optimum and sub-optimum quantum algorithms outperform the optimum classical algorithm in our setting. We have disregarded quantum processor initialization and network latency times because in a perfectly coupled quantum-classical hybrid computing system, this overhead would be negligible.

We are cautious not to claim that this algorithm in its current state using currently available quantum technology can outperform a state-of-the art graph coloring algorithm on top-of-the-line classical computing resources. Furthermore, we cannot draw any definitive conclusions for graphs of size $n > 60$ due to being unable to run larger graph sizes on the quantum annealer. The results of this experiment suggest potential quantum advantage and higher success probability using a quantum annealer as an independent set sampler as a part of a hybrid quantum-classical graph coloring algorithm.

\section{Conclusions}
\label{conclusions}

We have demonstrated an improvement in heuristic graph coloring computation time by substituting a quantum for a classical implementation of the maximum independent set solving subroutine. This result is significant not only because it demonstrates enhanced performance of the quantum solution, but because it includes time components such as quantum hardware readout and thermalization delay that are often left out of studies which address only core annealing time. The classical components of both algorithms were run on the same desktop workstation, providing a common environment for comparison. Taking all that into consideration, we view the result of faster heuristic graph coloring times over our problem set as extremely promising. It is an example of an application oriented problem where we have carefully chosen and mapped an appropriate subproblem to the quantum annealer, with demonstrably positive effects. 

The clear next step here would be to try larger problems on more advanced generations of quantum annealing hardware. We used the 1098 qubit DW2X annealing chip for this study to address graphs of up to 60 nodes. Researchers with access to larger DW2000Q or the Advantage platform would be able to extend the scaling graphs to larger problem sizes. In such a context, it would be possible to quantify the potential quantum advantage we observe with the small problems in this study and project scaling for both classical and quantum heuristic algorithms.

Another important extension of the work presented here would be testing with problems derived from real world applications. Logistics and wireless frequency allocation are both prime areas for problem generation. The constructed problems studied here are interesting because they sit at a transition in hardness of solution. Looking at real ensembles of instances, however, may point the way to areas where quantum annealing could be more advantageous than we have observed.

 \bibliographystyle{unsrt}
 \bibliography{quantum_gc_refs}

\end{document}